\documentclass[aps,prc,groupedaddress,twocolumn,showpacs,amsmath,amssymb]{revtex4}
\usepackage{graphics}
\usepackage{dcolumn}
\usepackage{longtable}

\begin{document}

\title{Low and medium energy deuteron-induced reactions on $^{63,65}$Cu nuclei}

\author{E. \v Sime\v ckov\'a} \email{simeckova@ujf.cas.cz}
\author{P. B\'em} 
\author{M. Honusek}
\author{M. \v Stef\'anik}
\affiliation{Euratom/IPP.CR Fusion Association, Nuclear Physics Institute, 25068 \v Re\v z, Czech Republic}

\author{U. Fischer}
\author{S.P. Simakov}
\affiliation{Euratom/FZK Fusion Association, Karlsruhe Institute of Technology (KIT), Hermann-von-Helmholtz-Platz, 1, 76344 Eggenstein-Leopoldshafen, Germany}

\author{R.A. Forrest}
\affiliation{Nuclear Data Section, International Atomic Energy Agency, A-1400 Vienna, Austria}

\author{A.J. Koning}
\affiliation{Nuclear Research and Consultancy Group, P.O. Box 25, NL-1755 ZG Petten, The Netherlands}

\author{J.-C. Sublet}
\affiliation{Euratom/UKAEA Fusion Association, Culham Centre for Fusion Energy, Abingdon OX14 3DB, United Kingdom}

\author{M.~Avrigeanu} \email{mavrig@ifin.nipne.ro}
\author{F.L. Roman}
\author{V.~Avrigeanu}
\affiliation{ ``Horia Hulubei'' National Institute for Physics and Nuclear Engineering, P.O. Box MG-6, 077125 Bucharest-Magurele, Romania}
	
\date{\today}

\begin{abstract}
The activation cross sections of $(d,p)$, $(d,2n)$, $(d,3n)$, and $(d,2p)$ reactions on $^{63,65}$Cu were measured in the energy range from 4 to 20 MeV using the stacked-foils technique. Then, following the available elastic-scattering data analysis that provided the optical potential for reaction cross sections calculations, an increased effort has been devoted to the breakup mechanism, the direct reaction stripping, and the pre-equilibrium and compound-nucleus cross section calculations, corrected for the breakup and stripping decrease of the total reaction cross section. The overall agreement between the measured and calculated deuteron activation cross sections proves the correctness of the nuclear mechanisms account, next to the simultaneous analysis of the elastic-scattering and reaction data.
\end{abstract}


\maketitle

\section{Introduction}
\label{Sec1}
Among the projects of powerful neutron sources for nuclear energy generation, the International Fusion Material Irradiation Facility (IFMIF) requests high accuracy deuteron evaluated nuclear data for the assessment of induced radioactivity of the accelerator components, targets and beam stoppers. The IFMIF facility needs such data for estimation of the potential radiation hazards from the accelerating cavities and beam transport elements (Al, Fe, Cr, Cu, Nb) and metal and gaseous impurities of the Li loop (Be, C, O, N, Na, K, S, Ca, Fe, Cr, Ni) in the energy range from the threshold up to 40 MeV. However, it is known that the actual experimental and evaluated data for deuteron-induced reactions are less extensive and accurate than for neutrons, so that further measurements and improved model calculations are needed. 

The weak binding energy of the deuteron, $B_d$=2.224 MeV, is responsible for the high complexity of the interaction process that involves a variety of reactions initiated by the neutrons and protons coming from deuteron breakup. Such a wide diversity of nuclear reactions initiated by deuteron interaction with nuclei has hampered so far the comprehensive analysis involving large A-range of targets and incident-energy domain. The difficulties to interpret the deuteron-induced reaction data in terms of the usual reaction mechanism models have recently been investigated \cite{bem08,ma09,bem09,ma10a,ma10b,es10}, looking for a consistent way to also include the breakup contribution within the activation cross section calculations. Second, the total reaction cross sections are less accurately described since, unlike the nucleon case, there are no global optical model potentials (OMP) which describe the scattering data over a wide range of nuclei and energies sufficiently well. Therefore, a simultaneous analysis of the deuteron elastic scattering \cite{ma09} and induced activation \cite{bem09}, which appears essential for the IFMIF engineering design, is extended in the present work for the $^{63,65}$Cu target nuclei. 

\section{Measurements} \label{Sec2}
\subsection{Samples and irradiations}

\begin{table} [b]
\caption{\label{tab:dec} Half-lives, main gamma lines and their intensities of the measured isotopes \cite{chu99}.}
\begin{tabular}{cccc}\\
\hline \hline
Isotope   & T$_{1/2}$ & E$_{\gamma}$ & I$_{\gamma}$ \\
          &           & (keV)  & (\%)\\
\hline
$^{65}$Zn & 244.26 d  & 1115.5 & 50.6  \\
$^{64}$Cu & 12.7 h    & 1345.8 & 0.473 \\
$^{62}$Zn & 9.186 h   & 596.6  & 26    \\
          &           & 548.4  & 15.3  \\
$^{65}$Ni & 2.5172 h  & 1481.8 & 24    \\
$^{63}$Zn & 38.47 min & 669.6  & 8     \\
          &           & 962.1  & 6.5   \\
$^{66}$Cu & 5.12 min  & 1039.2 & 9     \\
\hline \hline
\end{tabular}
\end{table}

The variable energy NPI cyclotron provides protons and deuterons in energy range 11-37 MeV and 11-20 MeV, respectively. The irradiation was carried out using an external deuteron beam of the NPI cyclotron U-120M operating in the negative-ion mode of acceleration. From the stripping-foil extractor the beam was delivered to the reaction chamber through a beam line consisting of one dipole and two quadrupole magnets. 

The incident deuteron energy was determined by computational procedure based on measured trajectory (the frequency and actual extraction radius) of acceleration. This procedure was experimentally tested using the activation foil method and the surface-barrier-detector technique. The energy was determined with a resulting accuracy of 1.0\%, the FWHM spread of the incident beam up to 1.8\% was observed.
            
The activation cross sections induced by deuterons bombarding high purity natural aluminum and copper foils were measured by a stacked-foil technique. A collimated deuteron beam strikes the stack of foils in a Faraday-cup-like reaction chamber enabling the cooling of stacked foils without a lost of accuracy in the beam current and charge monitoring (10\%). To enlarge the number of energy bins in the measured excitation functions and to check the internal consistency of the measured data, the foils were stacked with different Al vs Cu sequences in two independent runs. The stock of eleven Al and eleven Cu foils placed alternately was bombarded by a deuteron beam of initial deuteron energy 19.95 MeV, with mean beam current 90 nA during an exposure time of 15 min. The initial deuteron energy, the beam current and the irradiation time for the second run were 20.18 MeV, 330 nA and 5 min, respectively. The aluminum cross section data were reported earlier \cite{bem09}.

To enable the measurement of the relatively short-living isotopes $^{63}$Zn (T$_{1/2}$=38.47 min) and $^{66}$Cu (T$_{1/2}$=5.12 min), three extra runs were carried out. The initial deuteron energy was 19.79, 20.09 and 19.79 MeV, the beam current was 0.24, 0.36 and 0.17 $\mu$A  and the irradiation time was 6, 5 and 9.3 min, respectively.
The thickness of the high purity natural Cu and Al foils (purity of 99.99\%, Goodfellow product) was 25 and 50 $\mu$m, respectively. Foils were weighted (with a 2\% uncertainty) to avoid the relatively large uncertainties in the foil thickness declared by the producer. The mean deuteron energy and energy thickness were determined using the SRIM 2003 code \cite{zieg03}. The overall thickness of the available 22 foil stacks covers the excitation-curve range from 20 to full beam stop.

In preliminary reports \cite{bem08,es10}, different initial energies were reported due to errors in the orbit calculation of the cyclotron operated in the negative-ion mode of acceleration. In the present report, corrected energy values 19.95 and 20.18 MeV for the first two runs were established and the relevant energies and energy thicknesses of each foil were recalculated.       

\begin{table*} [htb] 
\caption{\label{tab:xsec}Measured reaction cross sections (mb) for deuterons incident on the $^{nat,63,65}$Cu nuclei. The mean deuteron energy and resolution due to the thickness and straggling of each foil are shown. Statistical uncertainties are given in parentheses for the cross sections in units of the last digit.}
\begin{tabular}{ccccccc}
\hline \hline
Energy& \multicolumn{6}{c}{Reaction}\\
\cline{2-7}
 (MeV) & $^{nat}$Cu$(d,x)^{64}$Cu & $^{63}$Cu$(d,2n)^{63}$Zn & $^{63}$Cu$(d,3n)^{62}$Zn & $^{65}$Cu$(d,p)^{66}$Cu & $^{65}$Cu$(d,2n)^{65}$Zn & $^{65}$Cu$(d,2p)^{65}$Ni \\
\hline
 1.49 (149)& 13.3 (16) \\
 4.25 (86) & 85.9 (91) &            &            &            & 9.7 (12)\\ 
 4.56 (79) &           & 1.70 (19)  &            & 123 (14) \\
 5.29 (75) & 195 (23)  &            &            &            &  92 (11)\\
 5.96 (67) &           & 1.58 (17)  &            & 282 (33) \\
 6.00 (66) &           &            &            & 258 (32) \\
 6.92 (63) & 214 (24)  &            &            &            & 243 (25) \\
 7.13 (60) &           & 8.29 (92)  &            & 322 (36) \\
 7.69 (59) & 253 (33)  &            &            &            & 445 (53) \\
 8.21 (54) &           & 42.0	(48)  &            & 290 (37) \\
 9.00 (53) & 216 (28)  &            &            &            & 531 (55)\\
 9.09 (50) &           & 103 (12)   &            & 284 (37) \\
 9.65 (51) & 218 (26)  &            &            &            & 675 (81) \\
 9.91 (47) &           & 155 (18)   &            & 261 (32) \\ 
10.07 (47) &           & 165 (23)   &            & 239 (34) \\
10.78 (47) & 182 (23)  &            &            &            & 683 (80) \\
11.36 (45) & 188 (25)  &            &            &            & 862 (102) & 0.101 (29)\\
11.55 (42) &           & 248 (30)   &            & 207 (24) \\
12.25 (41) &           & 281 (36)   &            & 196 (23) \\
12.37 (43) & 161 (17)  &            &            &            & 795 (83) \\
12.39 (41) &           & 269 (52)   &            & 179 (24) \\
12.90 (41) & 157 (21)  &            &            &            & 880 (104)& 0.200 (42)\\
13.69 (39) &           & 327 (41)   &            & 165 (20) \\
13.82 (39) & 147 (16)  &            &            &            & 940 (95) & 0.355 (73)\\
14.31 (37) & 134 (16)  & 355 (46)   &            & 166 (25)   & 909 (109)& 0.435 (59)\\
15.02 (36) &           & 353 (83)   &            & 136 (19) \\
15.17 (37) & 133 (15)  &            &            &            & 949 (95) & 0.502 (64)\\
15.62 (35) &           & 367 (53)   &            & 138 (19) \\
15.63 (36) & 114 (14)  &            &            &            & 911 (107)& 0.632 (94)\\
16.17 (34) &           & 380 (47)\\
16.44 (35) & 117 (12)  &            & 0.018 (4)  &            & 901 (91) & 0.750 (83)\\
16.87 (34) &  93 (11)  &            & 0.067 (36) &            & 851 (100)& 0.70 (11)\\
17.37 (33) &           &            &            & 109 (17) \\
17.38 (32) &           & 374 (55)   &            & 111 (14) \\
17.64 (33) & 103 (11)  &            & 0.253 (28) &            & 816 (82) & 0.92(10)\\
17.89 (32) &           & 382 (54)   &            & 103 (13) \\
18.05 (33) &  91 (11)  &            & 0.645 (89) &            & 817 (98) & 1.09 (17)\\
18.78 (32) & 102 (12)  &            & 1.44 (15)  &            & 790 (82) & 1.32 (18)\\
19.01 (30) &           & 366 (66)   &            & 91.6 (122)\\
19.17 (31) &  84 (11)  &            & 2.28 (28)  &            & 730 (88) & 1.27 (18)\\
19.49 (30) &           & 350 (59)   &            & 89.7 (105)\\
19.49 (29) &           & 316 (79)\\
19.88 (30) &  97 (12)  &            & 3.72 (38)  &            & 694 (71) & 1.47 (21)\\
\hline \hline
\end{tabular}
\end{table*}

\subsection{Calculation of cross sections and their errors}

The gamma-rays from the irradiated foils were measured repeatedly by two calibrated HPGe detectors of 23 and 50\% efficiency and of FWHM 1.8 keV at 1.3 MeV. To provide reliable corrections for the decay, the beam-current recorder and $\gamma$-ray spectrometer were synchronized in time. Activated isotopes were identified (see Table~\ref{tab:dec}) using nuclear decay data from Ref. \cite{chu99}. The measurements with different cooling times lasted up to 100 days after irradiation. By analyzing the spectra, the resulting activities at the end of irradiation were obtained. The uncertainty of 3\% includes statistical errors and the uncertainty of the detector-efficiency calibration.

In the case of short-lived isotope measurements, the irradiated Cu foils were immediately measured by the HPGe detector with 50\% efficiency. To reduce the dead time rate caused by the strong annihilation peak accompanying $\beta^+$-decay, the observed Cu foil was situated within two iron slides of 1 mm thickness and a lead plate of 10 mm thickness was placed between the measured foil and the HPGe detector. The detector efficiency was recalibrated according to the experimental conditions using a calibrated $^{152}$Eu radioactive source with an uncertainty of detector efficiency of 5\%.

The experimental cross sections, given in Table~\ref{tab:xsec}, are shown in Fig.~\ref{fig:ompx} and compared with previously measured data \cite{jlg63,cbf70,ho71,st06,ko07,fwp66,st01,mn06,nb63}. Their systematic errors are composed of current uncertainty (10\%), uncertainty of foil thickness (2\%) and uncertainty of detector efficiency determination (2\% and 5\%, respectively). The mean statistical error in activity determination was 2\%. Uncertainty of initial energy determination was 1\%, energy spread of incident beam up to 1.8\%.  Only energy thicknesses are shown in Fig.~\ref{fig:ompx}.  

$^{nat}$Cu$(d,x)^{64}$Cu. As the natural copper has two stable isotopes, the generation of $^{64}$Cu by irradiation of natural copper may proceed via three contribution reactions: $^{63}$Cu$(d,p)$ (with the threshold E$_{th}$=0 MeV), $^{65}$Cu$(d,t)$ (E$_{th}$=3.767 MeV), and $^{65}$Cu$(d,2np)$ (E$_{th}$=12.512 MeV). Therefore, even for deuteron energies below 20 MeV, we have to take into account all contributions and to report the measured data to the natural copper as shown in Fig.~\ref{fig:ompx}(a). 

$^{65}$Cu$(d,p)^{66}$Cu is the only possible reaction generating $^{66}$Cu isotope by deuteron irradiation on natural copper. These data are shown in Fig.~\ref{fig:ompx}(b). 

$^{63}$Cu$(d,2n)^{63}$Zn. There is only one possible way to generate $^{63}$Zn in the energy region up to 20 MeV, the experimental excitation function is shown in Fig.~\ref{fig:ompx}(c). 

$^{65}$Cu$(d,2n)^{65}$Zn. The $^{65}$Zn isotope can originate in the $^{65}$Cu$(d,2n)$ reaction, and also in the  $^{63}$Cu$(d,\gamma)$ reaction. The radioactive capture reaction cross sections are known to be very small (of the order of a few hundred $\mu$b or less). Hence, the $^{63}$Cu$(d,\gamma)$ reaction would not be expected to contribute appreciably to the measured yield of $^{65}$Zn [see Fig.~\ref{fig:ompx}(d)]. 

$^{63}$Cu$(d,3n)^{62}$Zn is the only possible reaction to generate $^{62}$Zn in energy region up 20 MeV. The cross sections are shown in Fig.~\ref{fig:ompx}(e).

$^{65}$Cu$(d,2p)^{65}$Ni reaction is the only possible way for generation of the $^{65}$Ni. Our cross section data (see Fig.~\ref{fig:ompx}(f)) are the first experimental values except the one value around the deuteron energy of 19 MeV.	

Overall, the present experimental data are in satisfactory agreement with the previous measurements \cite{jlg63,cbf70,ho71,st06,ko07,fwp66,st01,mn06,nb63} within one standard deviation except for the oldest data of Ref. \cite{jlg63} and several data of Ref. \cite{cbf70}.

\section{Energy--dependent optical potential}
\label{Sec3}

The description of the deuteron-nucleus interaction represents an important test for both the quality of reaction mechanism models and the evaluation of nuclear data. The simultaneous analysis of the deuteron elastic scattering and induced activation should really involve a consistent input of nuclear model calculations \cite{bem09,ma10a,ma10b}, a prime interest for the optical model potential (OMP) parameters being motivated by their further use in the analysis of all deuteron interaction cross sections.

\begin{figure} [t]
\resizebox{0.99\columnwidth}{!}{\includegraphics{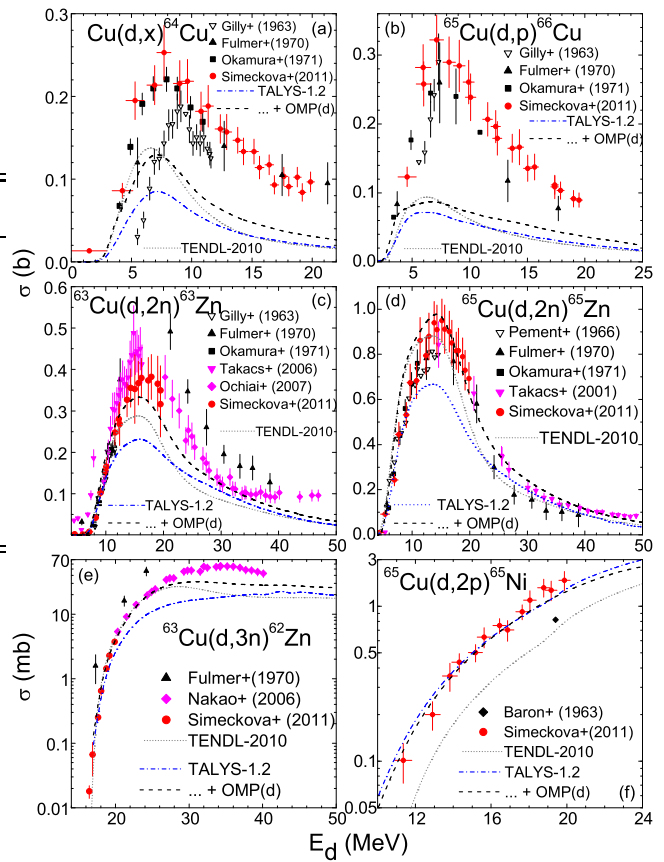}}
\caption{\label{fig:ompx}(Color online) Comparison of previous measured data \cite{jlg63,cbf70,ho71,st06,ko07,fwp66,st01,mn06,nb63} and within this work (full circles), the evaluated data within the TENDL-2010 library \cite{TENDL} (dotted curves), and calculated results obtained with TALYS-1.2 code using either the whole default input (dash-dotted) or the replacement of default deuteron OMP by that of the present work (dashed), for the deuteron interaction with $^{nat,63,65}$Cu.}
\end{figure}

\begin{table} [t]
\caption{\label{tab:omp} The parameters of the deuteron optical potential for the $^{63,65}$Cu target nuclei. A star used as superscript follows the parameters which were changed with respect to the optical potential of Daehnick et al. \cite{dah}.}
\begin{tabular}{lclc}\\
\hline \hline
\hspace*{0.4in}Potential depths &\hspace*{0.01in} &\multicolumn{2}{c}{Geometry parameters}\\
\hspace*{0.65in}     (MeV)                     & & \hspace*{0.75in}(fm)\\
\hline
V$_R$=88.5+0.88Z/A$^{1/3}$-0.26E               & & r$_R$=1.17\\
                                               & & a$_R^*$=0.734, &E$<$14.5  \\
                                               & & \hspace*{0.145in}=0.709+0.0017E, &E$>$14.5\\
W$_V^*$=-0.014+0.0948E                         & & r$_V$=1.325 \\
                                               & & a$_V$=0.810 \\
W$_D^*$=13.6,          \hspace*{0.56in} E$<$12 & & r$_D$=1.325 \\
\hspace*{0.22in}=15.27-0.142E, 12$<$E$<$35     & & a$_D^*$=0.770,& E$<$12 \\
\hspace*{0.22in}=10.3, \hspace*{0.56in} E$>$35 & & \hspace*{0.15in}=0.583+0.0156E,& 12$<$E$<$14.5\\
                                               & & \hspace*{0.15in}=0.810,        & E$>$14.5     \\
V$_{SO}$=7.33-0.029E                           & & r$_{SO}$=1.07\\
                                               & & a$_{SO}$=0.66\\
\hline \hline
\end{tabular}
\end{table}

\begin{figure} [b]
\resizebox{0.99\columnwidth}{!}{\includegraphics{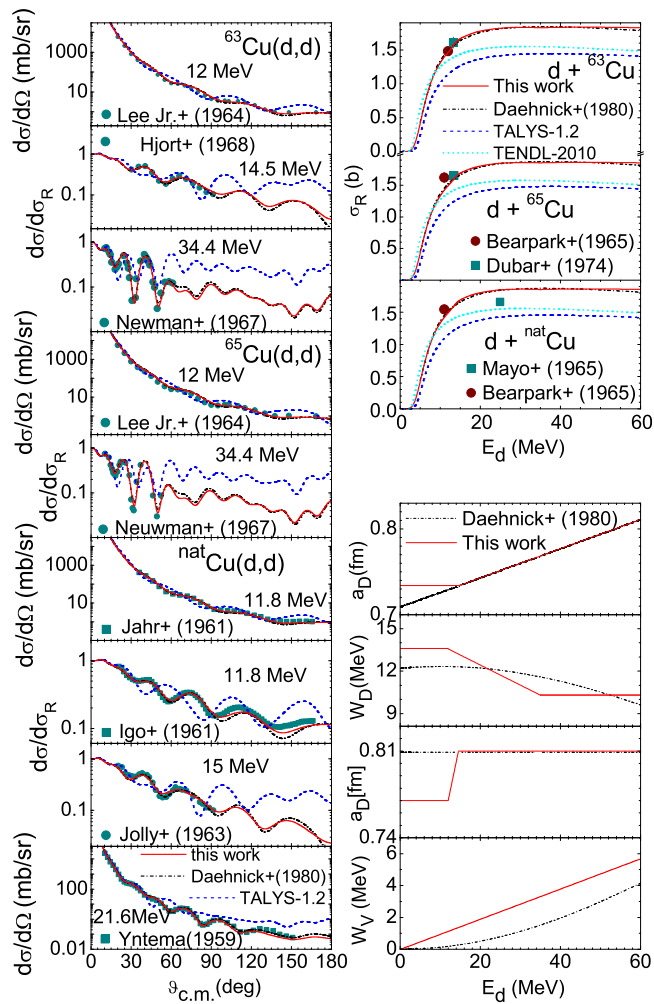}}
\caption{\label{fig:omp}(Color online) Comparison of (left) measured \cite{EXFOR_el} deuteron elastic-scattering angular distributions for $^{63,65,nat}$Cu and calculated values obtained by using the OMP parameters given in Table~\ref{tab:omp} (solid curves), the global OMP \cite{dah} (dot-dashed curves), and TALYS \cite{TALYS} default option (dashed curves), and (right) measured \cite{EXFOR-TCS} reaction cross sections for deuterons incident on $^{63,65,nat}$Cu and calculated values obtained by using the same above-mentioned potentials as well as the evaluated data within the TENDL-2010 library \cite{TENDL} (dotted curve). Comparison of particular parameters given in Table~\ref{tab:omp} (solid curves) and of Daehnick {\it et al.} \cite{dah} OMP (dot-dashed lines) are also shown in the lower part of the right hand side.}
\end{figure}

Unfortunately, the few measurements of angular distributions of elastic scattered deuterons on $^{63,65}$Cu \cite{EXFOR_el} do not allow an extended OMP analysis. However, while previous OMP analyses on $^{6,7}$Li \cite{ma05,ma06a}, $^{27}$Al \cite{ma09,bem09}, $^{54,56,58,nat}$Fe \cite{ma08}, $^{59}$Co and $^{93}$Nb \cite{ma10c} show that no global OMP describes sufficiently well the elastic scattering data in the energy range up to 20 MeV, but the few parameters adjustment (Table~\ref{tab:omp}) of the Daehnick et al. \cite{dah,RIPL3} OMP led to a good description of the data for the $^{63,65,nat}$Cu target nuclei. The comparison of the experimental elastic-scattering angular distributions for $^{63,65,nat}$Cu \cite{EXFOR_el} and the calculated values obtained by using the presently adjusted OMP parameters, the global optical potential \cite{dah}, and the widely-used TALYS code \cite{TALYS} default option based on the Watanabe folding approach \cite{wat58} (dashed curves) are shown in Fig.~\ref{fig:omp}.  At the same time the measured reaction cross sections for deuterons incident on the $^{63,65}$Cu isotopes and natural Cu \cite{EXFOR-TCS} are compared in Fig.~\ref{fig:omp}  with the calculated values obtained by using the same potentials as well as the evaluated data within the TENDL-2010 library \cite{TENDL}. One may note that the last two calculated excitation functions underestimate the measured values by at least 20\%. Finally, the present real-potential diffusibility, imaginary surface-potential depth and diffusibility, and imaginary volume-potential depth are compared with the same OMP parameters of Daehnick et al. \cite{dah}.  The elastic-scattering cross section calculations have been performed using the computer code SCAT2 \cite{SCAT2}.  

The particular importance of the deuteron OMP for the activation cross section calculations can be seen in Fig.~\ref{fig:ompx} through the comparison between the experimental data and the calculated results obtained using both the present deuteron OMP and the corresponding TALYS default option, as well as the TENDL-2010 library \cite{TENDL} data. There are thus compared the calculated cross sections obtained using the default input of TALYS, i.e. the Watanabe folding approach \cite{wat58} for the deuteron OMP, with the results following the replacement of this OMP by the parameter set given in Table~\ref{tab:omp}. The differences are obvious while it can also be seen that this replacement still does not lead to a satisfactory description of the experimental activation data.  Improvements of the theoretical analysis by taking into account all reaction mechanisms involved in the interaction process are thus additionally needed.

\section{Deuteron breakup} 
\label{Sec4}

\subsection{Phenomenological approach} 
\label{Sec4.1}

The interaction of deuterons with the target nuclei proceeds largely through direct reaction (DR) processes, for deuteron energies below and around the Coulomb barrier, while with increasing incident energy reaction mechanisms like pre-equilibrium emission (PE) or evaporation from the fully equilibrated compound nucleus (CN) also become important. On the other hand, the breakup mechanism is responsible for the enhancement of a large variety of reactions along the whole incident-energy range and thus its contribution to the activation cross sections has to be explicitly taken into account \cite{bem09,ma10a,ma10b}. 

\begin{figure} [t]
\resizebox{0.7\columnwidth}{!}{\includegraphics{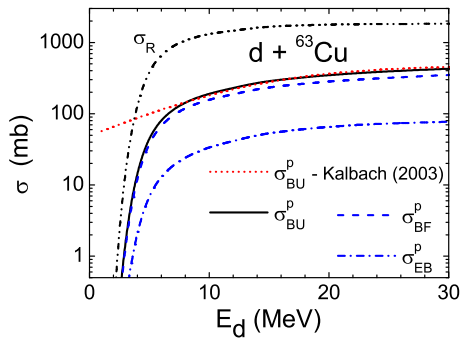}}
\caption{\label{fig:bu}(Color online) Energy dependence of the total (solid curve), elastic (dash-dotted) and inelastic (dot-dashed) proton-emission breakup cross sections given by Ref.  \cite{ma09} and total proton-emission breakup cross sections of Ref. \cite{kalb03} (dotted), for the deuteron interactions with $^{63}$Cu. The corresponding total reaction cross section is shown by dash-dot-dotted curve.}
\end{figure}

The physical picture of the deuteron breakup in the Coulomb and nuclear fields of the target nucleus considers two distinct chains, namely the elastic breakup (EB) in which the target nucleus remains in its ground state and none of the deuteron constituents interacts with it, and the inelastic breakup or breakup fusion (BF), where one of these deuteron constituents interacts with the target nucleus while the remaining one is detected. 
An empirical parametrization of the total proton-emission breakup fraction $f^{(p)}_{BU}$=$\sigma^{p}_{BU}$/$\sigma_R$ and the elastic-breakup fraction $f_{EB}$=$\sigma_{EB}$/$\sigma_R$ have previously been obtained \cite{ma09} on the basis of experimental systematics \cite{pamp78,wu79,klein81,mats82,must87}. Thus, proton-emission spectra and angular distributions from deuteron-induced reactions on nuclei from Al to Pb at incident energies from 15 to 80 MeV have been studied in this respect. However, an energy range of only 15-30 MeV has been available for the empirical elastic-breakup fraction $\sigma_{EB}$/$\sigma_R$ systematics \cite{klein81,must87}. Their dependence on the charge (Z), atomic number (A) of the target nucleus, and deuteron incident energy (E) was found to be \cite{ma09}:
\begin{eqnarray}\label{eq:1}
f^{(p/n)}_{BU}=0.087-0.0066 Z + 0.00163 ZA^{1/3}+ \:  \nonumber \\
 0.0017A^{1/3}E-0.000002 ZE^2, 
\end{eqnarray}
\begin{eqnarray}\label{eq:2}
f_{EB}=0.031-0.0028 Z + 0.00051 ZA^{1/3}+ \:  \nonumber \\
 0.0005A^{1/3}E-0.000001 ZE^2.
\end{eqnarray}
A comparison with the total proton- and neutron-emission breakup cross section parametrization of Kalbach \cite{kalb03}:
\begin{equation}\label{eq:3}
\sigma^{p/n}_{BU}=K_{d,(p,n)}\frac{(A^{1/3}+0.8)^2}{1+exp\frac{(13-E)}{6}}, \:\:  K_{d,p} = 21, \:\: K_{d,n} = 18.
\end{equation} 
shows that the former parametrization \cite{ma09} considers equal breakup fractions for proton and neutron emission, but also gives all breakup components, i.e. the proton-emission breakup total, elastic, and inelastic fraction $f^{(p/n)}_{BF}$=$f^{(p/n)}_{BU}$-$f_{EB}$. 
The energy dependence of the total, elastic, and inelastic proton-emission breakup cross sections following Ref.~\cite{ma09} as well as the total proton-emission breakup cross sections \cite{kalb03} for the deuteron interactions with the $^{63,65}$Cu nuclei are shown in Fig.~\ref{fig:bu}. It turns out that, for deuteron incident energies above ${\sim}$8 MeV, the predictions for the total proton-emission breakup cross sections given by both parameterizations are rather close. However, at the lowest energies the total proton-emission breakup cross section provided by the latter parametrization \cite{kalb03} become larger than the deuteron total reaction cross section. 

Concerning the energy dependence of the inelastic- and elastic-breakup components, the interest of the deuteron activation cross sections for incident energies up to 60 MeV has motivated an additional check of the elastic-breakup parameterization \cite{ma09} extension beyond the energies formerly considered for the derivation of its actual form. Actually, as it is shown in Fig.~\ref{fig:cdcc} for the $^{63}$Cu target nucleus, the elastic-breakup cross sections given by the empirical parameterization \cite{ma09} decrease with the incident energy beyond the energy range within which it was established. On the other hand, this trend is opposite to that of the total-breakup cross section. Thus, in the absence of any available experimental deuteron elastic-breakup cross section at incident energies above 30 MeV, the correctness of an eventual extrapolation should be checked by comparison of the related predictions with results of an advanced theory such as the Continuum-Discretized Coupled-Channels (CDCC) method \cite{kamimura86,aust87}.

\subsection{Phenomenological EB versus CDCC formalism} 
\label{Sec4.2}
A detailed description of the CDCC formalism is available elsewhere \cite{kamimura86,aust87,piya99,moro06,moro09b,moro09c}, and hence only a brief description of the method is given in the following.

The breakup component is treated within the CDCC formalism as an inelastic excitation of the projectile due to the nuclear and Coulomb interactions with the target nucleus. Consideration of this excitation is performed through the coupling of the projectile unbound excited states in the solution of the scattering problem by means of the coupled channels approach. The deuteron scattering process is analyzed within a three-body model, comprising the two-body projectile and the target, by the model Hamiltonian \cite{aust87}:
\begin{equation}\label{eq:4}
H=K_r+K_R+V_{np}({\bf r})+U_n({\bf R}-{\bf r}/2)+U_p({\bf R}+{\bf r}/2).
\end{equation}
Here  $V_{np}$ is the interaction between the neutron and proton \cite{kamimura86}, assumed to have a Gaussian shape 
\begin{equation}\label{eq:5}
V(r)=-V_0 e^{-(r/r_0)^2} ,
\end{equation}
where  $V_0$=72.15 MeV and $r_0$=1.484 fm , were determined from the fit of the deuteron binding energy. 
The vector ${\bf r}$ is the proton-neutron relative coordinate, ${\bf R}$ is the coordinate of the center of mass of the $p$-$n$ pair relative to the target nucleus. $U_p$ and $U_n$ are the proton-target and  neutron-target interactions, respectively, usually taken as the central nuclear part of the proton and neutron phenomenological OMPs at half the deuteron incident energy, $E_d/2$. Adjusted Koning-Delaroche \cite{KD03} neutron and proton global OMPs, in order to obtain a suitable description of the deuteron elastic-scattering, have been used \cite{maCDCC10}. The operators $K_r$ and $K_R$ are the kinetic energies associated with ${\bf r}$ and ${\bf R}$.

A finite set of coupled equations is obtained by the introduction of the discretization procedure in which the continuum spectrum, truncated at a maximum  excitation energy ($E^{*}_\mathrm{max}$) and divided  into a set of  energy (or relative momentum) intervals, is represented by a finite and discrete set of square-integrable functions. Each \textit{bin}, is represented by a single square-integrable function,  calculated by averaging the scattering states for the $p$-$n$ relative motion  within the bin width. Moreover, the $p$-$n$ relative angular momentum is also restricted by considering only a limited number of partial waves, in order to deal with a finite set of coupled equations. Finally, the three-body scattering wave function is expanded  over the internal states of the deuteron as follows:
\begin{equation}\label{eq:6}
\left|\Psi(E)\right\rangle=\sum_{i=0}^N\left|\phi_i,\chi_i\right\rangle,
\end{equation}
where $\left|\phi_0\right\rangle$ is the ground-state wave function and $\phi_i$ ($i\neq0$) are the averaged (within each bin) continuum wave functions. The radial functions $\chi_{i}(\bf R)$ describe the projectile-target relative motion for the elastic $(i=0)$ and breakup $(i\neq0)$ components. Continuum states with  orbital angular momentum $\ell=0$, 1 and 2 for the $p$-$n$ relative motion were considered. The proton and neutron intrinsic spins were ignored for simplicity within calculations that were performed with the coupled-channels code {\sc FRESCO} \cite{thompson88}.

\begin{figure} [t]
\resizebox{0.7\columnwidth}{!}{\includegraphics{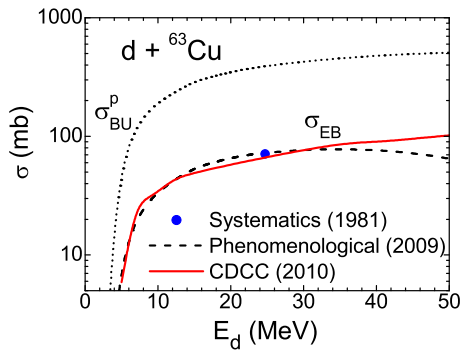}}
\caption{\label{fig:cdcc}(Color online) Energy dependence of the phenomenological \cite{ma09} (dashed curve) and CDCC (solid line) elastic-breakup cross sections for deuteron interactions with the $^{63}$Cu nucleus \cite{maCDCC10}. The solid circle corresponds to Kleinfeller systematics \cite{klein81} while there is also shown the total proton-emission breakup cross section (dotted curve).}
\end{figure}

The energy dependence of the elastic-breakup cross sections provided by the excitation of the continuum spectrum, in the case of the deuteron interaction with $^{63}$Cu target nucleus, is compared with the prediction of empirical systematics \cite{ma09} in Fig.~\ref{fig:cdcc}. The elastic-breakup cross sections corresponding to the Kleinfeller \textit{et el.} systematics  (Table 3 of Ref.~\cite{klein81}) are also shown. The agreement of the CDCC elastic-breakup cross sections and the latter systematics can be considered as  validation of the present advanced model approach. Moreover, the comparison shown in Fig.~\ref{fig:cdcc} points out that the CDCC calculations lead to elastic-breakup cross sections which follow the total-breakup cross section behavior as well as that of the reaction cross section shown in Fig.~\ref{fig:bu}. Therefore the present analysis makes clear that the empirical parameterization extension for the elastic-breakup cross sections beyond the energies considered in this respect should be done with caution \cite{maCDCC10}.  The CDCC method thus provides useful initial guidance for the assessment of these extrapolation accuracies and may help to improve the existing phenomenological approaches. However, additional experimental deuteron interaction data, like elastic-scattering angular distributions and inclusive neutron and proton spectra, are needed in order to validate the parameters involved in the CDCC and complete the systematics of the elastic- and total-breakup cross sections over enlarged energy and target mass domains.

\subsection{Inelastic-breakup enhancement of the deuteron activation cross sections} 
\label{Sec4.3}

Overall, the deuteron total-breakup cross section should be subtracted from the total reaction cross section that is shared among different statistical--emission channels. On the other hand, the inelastic-breakup processes, where one of deuteron constituents interacts with the target leading to a secondary composite nucleus, bring contributions to different reaction channels. The secondary--chance emission of particles from the original d-target interaction is therefore especially enhanced. Thus, the absorbed proton or neutron following the breakup emission of a neutron or proton, respectively, contributes to the enhancement of the corresponding $(d,xn)$ and $(d,xp)$ reaction cross sections. In order to calculate this breakup enhancement for, e.g., the $(d,xn)$ reaction cross sections, firstly the inelastic-breakup cross sections $\sigma_{BF}^{n}$ were obtained in the present work by  subtracting also the CDCC elastic-breakup cross sections from the phenomenological total-breakup cross sections given by Eq. (1). Next, they have been \cite{bem09,ma10a,ma10b,es10} multiplied by the ratios $\sigma_{(p,x)}$/$\sigma_R$ corresponding to the above-mentioned reactions of the absorbed proton with the target nucleus, where $\sigma_R$ is the proton reaction cross section and $x$ stands for  $n$ or $2n$ outgoing channels \cite{bem09}. These ratios have been expressed as a function of the deuteron incident energy using the Kalbach \cite{kalb07} formula for the center-of-mass system centroid energy of the deuteron-breakup peak energies of the emitted constituents:
\begin{equation}\label{eq:7}
\epsilon_{n,p}=\frac{1}{2}(\epsilon_d-B_d\mp \frac{1.44 Z}{1.5A^{1/3}+3.1}).
\end{equation}
In a similar way have been obtained also the inelastic-breakup contributions to the $(d,p)$ and $(d,2p)$ activation cross section due to the neutrons absorbed in further interactions with the target nucleus, i.e. by the $(n,\gamma)$, and $(n,p)$ reactions, respectively. The only difference has consisted in replacing the above-mentioned ratios $\sigma_{(p,x)}$/$\sigma_R$ by the ratios $\sigma_{(n,x)}$/$\sigma_{non}$, where the non-elastic cross section $\sigma_{non}$ plays the same role for neutrons as $\sigma_R$ for protons. 

\begin{figure} [t]
\resizebox{0.8\columnwidth}{!}{\includegraphics{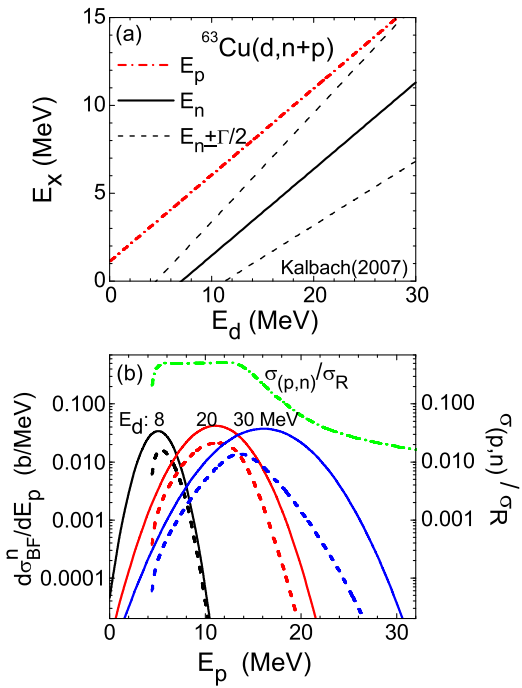}}
\caption{\label{fig:fracs}(Color online) (a) The centroid $E_x$ of assumed Gaussian line shape \cite{kalb07} for deuteron breakup-peak energies of emitted neutrons (solid) and protons (dash-dotted), and the corresponding $E_n\pm\Gamma/2$ values (dashed) calculated for deuterons interacting with $^{63}$Cu. (b) The convolution of the cross section ratio $\sigma_{(p,n)}$/$\sigma_R$ for the target nucleus $^{63}$Cu (dash-dotted curve) with the Gaussian line shape of the deuteron-breakup peak energies of the corresponding emitted protons, for deuterons with energies of 8, 20 and 30 MeV (solid curves, with the incident energy noted above their maxima), and the convolution results at each deuteron energy (dashed curves).}
\end{figure}

However, the assumed Gaussian line shape of the deuteron-breakup peak energies of the emitted constituents, that are also showed in Fig.~\ref{fig:fracs}(a) for neutrons, have quite large widths. Since the broad approximation of this method adopted previously \cite{bem09} for estimation of the breakup enhancement, a better estimation is considered in the present work.  It consists in a convolution of either the ratio  $\sigma_{(n,x)}$/$\sigma_{non}$ or $\sigma_{(p,x)}$/$\sigma_R$ with the Gaussian line shape of the deuteron-breakup peak energies of the corresponding emitted constituent, for a given deuteron incident energy. The cases of deuterons with energies of 8, 20 and 30 MeV are shown in Fig.~\ref{fig:fracs}(b) together with the cross section ratio $\sigma_{(p,n)}$/$\sigma_R$. There are also shown the convolution results at each of these  energies, while their area corresponds to the inelastic-breakup enhancement of the $(d,2n)$ reaction cross sections at these energies. These results are more physical, with a realistic incident-energy dependence except only for the case involving the higher-emission energy side of the Gaussian line shape of the deuteron-breakup peak energies. This happens for, e.g., deuterons with energies lower than 8 MeV, shown for the $(d,2n)$ reaction case in Fig.~\ref{fig:fracs}(b). Since this side of the Gaussian line shape could be narrower \cite{kalb07}, different widths for the two halves of the Gaussian distribution should be eventually adopted, while otherwise some overestimation may result from using a single width. However, the corresponding $(d,2n)$ reaction cross sections at these energies are just above the reaction threshold so that we have taken into account the inelastic-breakup enhancement of the $(d,2n)$ reaction cross sections only above these energies.

\section{One-nucleon transfer reactions}
\label{Sec5}

Apart from the breakup contributions to deuteron interactions, the direct reaction mechanisms like stripping and pick-up have to be properly considered in order to describe the low energy side of the $(d,p)$, $(d,n)$ and $(d,t)$ excitation functions \cite{ma09,bem09,ma10a,ma10b,es10,ma10c}. 
In the present work the DR contribution to the $(d,p)$ and $(d,t)$ reaction reaction cross sections, through population of the low-lying discrete levels of $^{64,66}$Cu residual nuclei, was calculated using the code FRESCO \cite{thompson88} based on the Coupled--Reaction Channels (CRC) method. The post form distorted-wave transition amplitude with finite-range interaction has been chosen.

\subsection{The one-nucleon stripping}
\label{Sec5.1}

The $(d,p)$ reaction has been in a large extent of critical importance for the study of nuclear structure. Actually, the spectroscopic factors extracted from the analysis of experimental angular distributions of the corresponding emitted protons did contribute to the validation of the nuclear shell model, considering that the neutron from the deuteron is transferred to a single-particle state of the residual nucleus. Consequently, the rich systematics of the achieved experimental spectroscopic factors makes possible the calculation of almost total stripping cross--section contribution to the deuteron activation cross section.  

The above-mentioned deuteron phenomenological OMP parameter set (Table~\ref{tab:omp}) has been used for the incident channel, while the Koning--Delaroche \cite{KD03} OMP global parameters have been used for protons interactions with the residual nuclei. The $n$-$p$ interaction in the deuteron has been described with the potential $V_{np}$ given by Eq. ($\ref{eq:5}$) and the neutron bound states were generated in a Woods--Saxon real potential with the global values of a reduced radius of 1.25 fm and diffuseness of 0.65 fm, while its depth has been adjusted to reproduce the nucleon binding energies in the residual nuclei. 

The present calculations of the single-neutron stripping $(d,p)$ reaction cross section have involved transitions to 104 final states of the odd-odd residual nucleus $^{64}$Cu and to 81 final states of the similar residual nucleus $^{66}$Cu. The spectroscopic factors that were obtained experimentally from proton angular-distribution measurements for these states up to $\sim$5 MeV, as given in Table II of Ref. \cite{park63}, Table I of Ref. \cite{park65}, and \cite{NDS108,NDS111}, have been considered in this respect. One may note that a lower number of final states, i.e. 63 for $^{64}$Cu and 52 for $^{66}$Cu, extending up to $\sim$3 MeV, were used within a preliminary stage of this work \cite{ma10a}, which made necessary an additional assumption concerning the DR contribution from the states at higher energies. The present increase of the final states taken into account for the DR contribution to the $(d,p)$ reaction cross sections makes a similar assumption no longer necessary. In spite of the corresponding results being rather close to the former ones \cite{ma10a}, an increased accuracy is now obtained for this activation component, even better than $\sim$5\%. 

The resulted stripping components of the $(d,p)$ reaction excitation functions are essential for the description of the experimental data shown in  Fig.~\ref{Fig_Cu635cs}. This statement is valid also concerning the maxima of these excitation functions at deuteron incident energies $E_{d}$$\sim$8 MeV. 

\subsection{The one-nucleon pick-up}
\label{Sec5.2}

The excitation function corresponding to $^{64}$Cu nucleus production from deuterons incident on a natural copper target includes contributions from both stripping $^{63}$Cu$(d,p)$ and  pick-up $^{65}$Cu$(d,t)$ direct reaction mechanisms. Actually, the lowest energy side of a $(d,t)$ excitation function, between its threshold and those of the $(d,dn)$ and $(d,p2n)$ reactions, can be described exclusively by the pick-up reactions as it is shown, e.g., in Fig. 3 of Ref. \cite{ma10c}. 

Therefore, we have used in the present work the above-mentioned deuteron phenomenological OMP to  describe the incident channel of the $(d,t)$ reaction, in a similar way to the stripping calculation, while the Becchetti and Greenlees \cite{RIPL3,BG} OMP has been used for the emitted tritons.

The $d$-$n$ interaction in triton has been described with a $^3$He potential \cite{3He-pot} of Woods-Saxon shape: 
\begin{equation}\label{eq:8}
V(r)=-V_0 \frac{1}{1+e^{(r-r_0)/a}} ,
\end{equation}
\noindent
where  the parameters $V_0$=77.71 MeV, $r_0$=1.008 fm and $a$=0.75 fm  were determined by a fit of the 5.50 MeV $^3$He binding energy (relatively close to 6.3 MeV corresponding to $^3$H). The transfered--neutron bound states were also generated in a Woods-Saxon real potential, with global reduced radius of 1.25 fm, diffuseness of 0.65 fm, and the depth adjusted to the nucleon binding energies in the residual nuclei. The experimental spectroscopic factors obtained by analysis of triton angular distributions related to the population of 14 discrete levels of the residual nucleus $^{64}$Cu \cite{NDS108} have been used for calculation of the pick-up transition amplitude. 

\begin{figure*} [htb]
\resizebox{1.5\columnwidth}{!}{\includegraphics{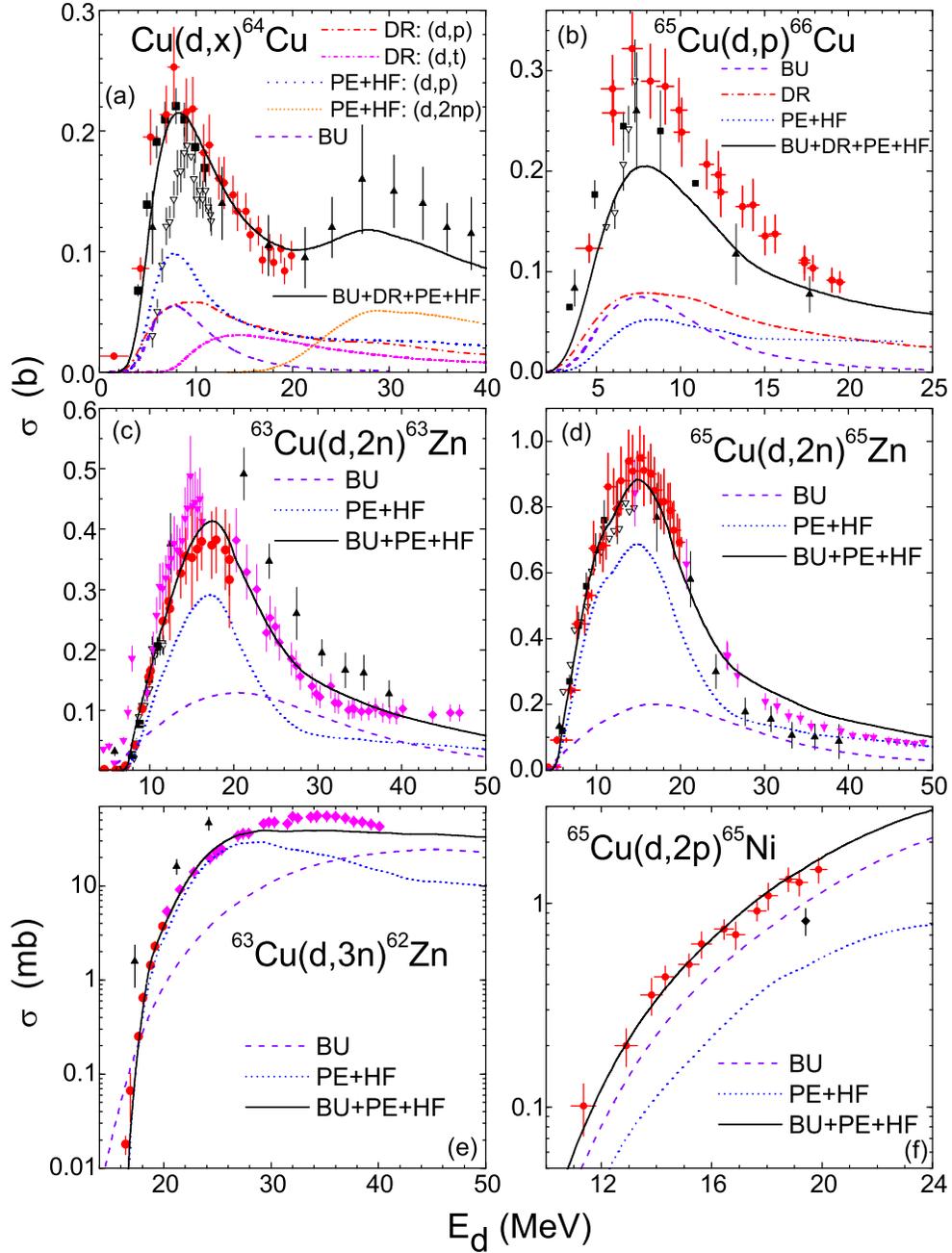}}
\caption{\label{Fig_Cu635cs}(Color online) Comparison of measured data already shown in Fig.~\ref{fig:ompx} and present analysis results (solid) taking into account the deuteron inelastic breakup (dashed), the DR (dash-dotted for $(d,p)$ reactions and short dash-dotted for $(d,t)$ reaction), and the PE+CN (dotted and short-dotted for $(d,2np)$ reaction) mechanism contributions to the deuteron interaction with $^{nat,63,65}$Cu target nuclei.}
\end{figure*}

The contribution thus obtained of the $^{65}$Cu$(d,t)$ reaction to the $^{64}$Cu production excitation function, corrected for the isotopic ratio $^{65}$Cu/$^{63}$Cu, is shown in Fig.~\ref{Fig_Cu635cs}(a). As expected \cite{ma10c}, the $(d,t)$ activation cross sections is the essential contribution among the processes induced by deuterons on the $^{65}$Cu isotope, within the natural copper target, at incident energies between 7-20 MeV. The statistical emission through the $(d,2np)$ and $(d,nd)$ reactions, also shown in Fig.~\ref{Fig_Cu635cs}(a), become significant at higher energies.

\section{Statistical particle emission}
\label{Sec6}

The reaction mechanisms such as pre-equilibrium emission (PE) or evaporation from the fully equilibrated compound nucleus (CN), become important when the incident energy is increased above the Coulomb barrier, when the interaction of deuterons with the target nuclei proceeds largely by DR processes. The related cross sections have been analyzed in this work by using the default model parameters (except for the deuteron OMP in  Table~\ref{tab:omp}) of the widely-used computer code TALYS as well as a local consistent parameter set developed in calculations with the PE+CN code STAPRE-H \cite{ma95} taking into account also the breakup and DR results discussed above. 

The local analysis results obviously have a higher accuracy while the global predictions may be useful for an understanding of unexpected differences between measured and calculated cross sections. The main assumptions and parameters involved in this work for the sets of global and local calculations have recently been described elsewhere \cite{va08}, only some points specific to the mass range A$\geq$60 are given here. A further note should concern the fact that similar input parameter sets and calculations have been used to obtain the breakup-enhancement due to one of deuteron constituents interacting with the target nucleus and leading to a secondary composite nucleus, with final contributions to different reaction channels as discussed in Sec.~\ref{Sec4}.

The deuteron phenomenological optical model parameter set given in Table~\ref{tab:omp} has been used for the incident channel. The nucleon optical potential of Koning and Delaroche \cite{KD03}, used by default in the TALYS code, has obviously been the first option. However, a basic point revealed by these authors is that their global potential does not reproduce the minimum around the neutron energy of 1--2 MeV for the total neutron cross sections of the $A$$\sim$60 nuclei. Following also their comment on the constant geometry parameters which may be responsible for this aspect, we have applied the SPRT method \cite{jpd76} for determination of the OMP parameters over a wide neutron energy range through analysis of the $s$- and $p$-wave neutron strength functions, the potential scattering radius $R'$ and the energy dependence of the total cross section $\sigma_T (E)$. The recent RIPL-3 recommendations \cite{RIPL3} for the low--energy neutron scattering properties and the available measured $\sigma_T$ data have been used in this respect, and we found that it is necessary to consider the energy dependence of the real potential geometry at lower energies given in Ref. \cite{va08}. 

These potentials were also used for the calculation of the collective inelastic scattering cross sections by means of the direct--interaction 
distorted--wave Born approximation (DWBA) method and a local version of the computer code DWUCK4 \cite{pdk84}. The weak coupling model was adopted for the odd nuclei $^{55}$Mn and $^{63,65}$Cu using the collective state parameters of Kalbach \cite{ck00}. Typical ratios of the direct inelastic scattering to the total reaction cross sections in the energy range from few to 60 MeV decrease from $\sim$11 to 5\%, for the $^{55}$Mn nucleus, and from $\sim$8 to 3\% for the Cu isotopes. 

The OMP of Koning and Delaroche was also considered for the calculation of proton transmission coefficients on the residual nuclei, i.e. the isotopes of Cu and Ni, while a previous trial of this potential concerned the proton reaction cross sections $\sigma_R$ \cite{rfc96}. Actually our local analysis involved the isotopes of Mn, Fe, Co, Ni, Cu and Zn, for lower energies important in statistical emission from excited nuclei. In order to obtain the agreement with the corresponding $\sigma_R$ data we have found it necessary to replace the constant real-potential diffusivity by the energy--dependent form $a_V$=0.463+0.01$E$ up to 20 MeV for $^{58}$Ni, where the energy $E$ is in MeV and the diffusivity is in fm. A final validation of both the original OMP and the additional energy--dependent $a_V$ has been obtained by analysis of the available $(p,\gamma$) and $(p,n)$ reaction data up to $E_p\sim$12 MeV on Ni isotopes while the other statistical model parameters are the same as in the rest of the present work. 

The  optical potential which is used in this work for calculation of the $\alpha$-particle transmission coefficients was established previously \cite{va94} for emitted $\alpha$-particles, and supported recently by semi--microscopic analysis for $A$$\sim$90 nuclei \cite{ma06b}. On the other hand, by comparison of the present calculations and measured data \cite{EXFOR} for the target nuclei $^{63,65}$Cu we found that the real well diffuseness $a_R$ of the above--mentioned global OMP should be changed to 0.67 fm. This reduction is rather similar to that found necessary for the target nuclei $^{59}$Co, and $^{58,60,62}$Ni \cite{vs04}. 
Moreover, a final remark concerns the fact that the same OMP parameter sets were employed within both the PE generalized \cite{ma95} Geometry-Dependent Hybrid (GDH) model \cite{mb83}, and the CN statistical model. 

The nuclear level densities were derived on the basis of the back-shifted Fermi gas (BSFG) formula \cite{hv88}, for the excitation energies below the neutron-binding energy, with small adjustments of the parameters $a$ and $\Delta$ \cite{va02} obtained by a fit of more recent experimental low-lying discrete levels \cite{ensdf} and $s$-wave nucleon resonance spacings $D_0$ \cite{RIPL3}. Above the neutron binding  we  took into account the washing out of shell effects within  the approach of Ignatyuk  et  al. \cite{avi75} and Junghans et al. \cite{arj98}, and using the method of Koning and Chadwick \cite{ajk97} for fixing the appropriate shell correction energy. A transition range from the BSFG formula description to the higher energy approach has been chosen between the neutron binding energy and the excitation  energy of 15 MeV, mainly in order to have a smooth connection. On the other hand, the spin distribution has been determined by a variable ratio $I/I_r$ of the nuclear moment of inertia to its rigid-body value, between  0.5 for  ground  states, 0.75 at the neutron binding energy, and 1 around  the excitation energy of 15 MeV. Concerning the particle-hole state density which for the PE description plays the same role as the nuclear-level density for statistical model calculations, a composite formula \cite{ma98} was used within the GDH model with no free parameters except for the $\alpha$-particle state density $g_{\alpha}$=$A/10.36$ MeV$^{-1}$ \cite{eg81}.

The modified energy--dependent Breit--Wigner (EDBW) model \cite{dgg79,mav87} was used for the electric dipole $\gamma$-ray strength functions $f_{E1}(E_{\gamma})$ of main importance for calculation of the $\gamma$-ray  transmission coefficients. The corresponding $f_{E1}(E_{\gamma})$ values have been checked within the calculations of capture cross sections of Mn and Cu isotopes in the neutron energy range from keV to 3--4 MeV, by using the OMP and nuclear level density parameters described above and global estimations \cite{chj77} of the $\gamma$-ray strength functions for multipoles $\lambda$$\le$3. Thus we found that the $f_{E1}(E_{\gamma})$ strength functions corresponding to the experimental \cite{RIPL3} average radiative widths $\Gamma_{\gamma0}^{exp}$ provide an accurate description of the capture data for the Cu isotopes. Finally, the accuracy of the $\gamma$-ray strength functions adopted in this work is also shown by the above--mentioned analysis of the $(p,\gamma)$ reaction cross sections.

Formally, no free parameter is involved for the PE description within the corresponding generalized GDH model except for $\alpha$-particle emission, the above-mentioned s.p.l.-density and the pre-formation probability $\varphi$ \cite{eg81} with a value of 0.2 used in the present work. However, a particular comment concerns the initial configuration of excited particles ($p$) and holes ($h$), for deuteron-induced reactions in the present case. Similar careful studies \cite{klein81,must87,pamp78,hiw87} pointed out that $3p$-$1h$ or $2p$-$1h$ may be a suitable choice for this configuration. Our calculations show that the latter one gives the best agreement between the measured and calculated reaction cross sections.

\section{Results and discussion} 
\label{Sec7}

The comparison of the measured and calculated $(d,p)$ reaction cross sections of $^{nat,63,65}$Cu is shown in Fig.~\ref{Fig_Cu635cs}, including the present experimental data already shown in Fig.~\ref{fig:ompx}, and the global and local analysis results. For the local analysis both components of the final activation are shown, i.e. the DR cross sections provided by the code FRESCO and the PE+CN contributions supplied by STAPRE-H. The latter is alone rather close to the TALYS predictions. The local approach has led to much better agreement with the present $(d,p)$ reaction data \cite{bem08,es10} especially due to the stripping DR contribution.

In order to obtain a complete description of the $(d,2p)$ reaction cross sections, we have started by taking into account also the neutrons which, following  the breakup proton emission, are absorbed in further interactions with the target nucleus. The cross section  $\sigma_{BF}^{p}$ \cite{ma08} has been considered in this respect (Fig.~\ref{fig:fracs}) as well as the corresponding fraction leading to the above-mentioned reactions. These fractions have been obtained by using the ratios of the most recently evaluated \cite{mbc06} $(n,p)$ and $(n,\alpha)$ reaction cross sections, respectively, to the neutron reaction cross sections provided by the neutron global OMP \cite{KD03}. A similar procedure has been followed in order to obtain the contribution to the $(d,2n)$ and $(d,3n)$ reaction cross sections due to the protons which, following the breakup neutron emission, are absorbed in further interactions with the target nucleus and described by the cross section $\sigma_{BF}^{n}$. The only difference in this case concerns the $(p,n)$ reaction cross sections in the incident energy range up to 30 MeV, which have been obtained by PE+CN calculations using the computer code STAPRE-H and the consistent local parameter set described above. All intermediary and ultimate reaction cross sections shown in Fig.~\ref{Fig_Cu635cs}(c-e) indicate that they may contribute up to 50\% of the activation cross sections for deuteron incident energies of $\sim$25 MeV.

The contribution due to the breakup proton, added to the PE+CN components provided by STAPRE-H, describe rather well the measured cross sections of the $(d,2p)$ reaction as shown in Fig.~\ref{Fig_Cu635cs}(f). Similarly, the breakup neutron emission plays the same role for the $(d,2n)$ reaction as shown in Fig.~\ref{Fig_Cu635cs}(c,d). Their weight is obviously increasing with the incident energy since all reactions involved, following the deuteron breakup, within the second step of these processes have negative Q-values.  

Finally, all activation data of deuteron-induced reactions on $^{63,65}$Cu have been properly described, making obvious the usefulness of the concurrent description of all reaction channels as well as the simultaneous analysis of the deuteron elastic scattering and induced activation. A particular underprediction has concerned however the $^{65}$Cu$(d,p)^{66}$Cu reaction cross sections, only their energy dependence being well described. A first comment may concern in this respect the fact that, although the TALYS and TENDL calculations do include a breakup component in all $(d,n)$ and $(d,p)$ reaction channels, the systematical relations for its strength does not show enough predictive power in this particular case. On the other hand, also the lower number of the known spectroscopic factors  corresponding to the discrete states of the odd-odd residual nucleus $^{66}$Cu, taken into account for the DR contributions, may explain this underprediction.

\section{Summary} 
\label{Sec8}

The cross section values for deuteron-induced reactions on natural Cu were determined for the reactions  $^{nat}$Cu$(d,x)^{64}$Cu, $^{65}$Cu$(d,p)^{66}$Cu,  $^{63}$Cu$(d,2n)^{63}$Zn, $^{65}$Cu$(d,2n)^{65}$Zn, $^{63}$Cu$(d,3n)^{62}$Zn and $^{65}$Cu$(d,2p)^{65}$Ni at deuteron energies up to 20 MeV. Resulting cross section data are in good agreement with the major part of previous reported experiments. 

Following a previous extended analysis of elastic-scattering, breakup and direct-reactions of deuterons on $^{63,65}$Cu, for energies from 3 to 60 MeV \cite{ma10a}, the pre-equilibrium and statistical emissions have been considered in the same energy range. The related cross sections have been analyzed by using the default model parameters (except for the deuteron OMP in  Table~\ref{tab:omp}) of the widely-used computer code TALYS as well as a local consistent parameter set developed in calculations with the PE+CN code STAPRE-H taking into account also the breakup and DR results formerly discussed. The local approach has led to much better agreement with the present $(d,p)$ reaction data especially due to the model calculation of the stripping DR contribution.

Consideration of the deuteron breakup plays a key role for the reaction channels adding a second emitted particle to the first one. Thus, in order to obtain a complete description of the $(d,p)$, $(d,2n)$, $(d,3n)$, and $(d,2p)$  reaction cross sections, we have taken into account also the neutrons which, following  the breakup proton emission, are absorbed in further interactions with the target nucleus. Finally, all deuteron-induced reactions on $^{63,65}$Cu,  including the present data measured at 20 MeV deuteron energy, have been properly described due to a simultaneous analysis of the elastic-scattering and reaction data. A similar analysis will be further considered for a systematical evaluation of the deuteron activation of other medium-mass nuclei.

\section*{Acknowledgments}
The authors are indebted to the operating crew of the U-120M cyclotron for their ready assistance. This work was partly supported by the MTI CR under contract No. 2A-1TP1/101, by the European Communities within the framework of the European Fusion Development Agreement under Contracts of Association between EURATOM and Forschungszentrum Karlsruhe, IPP.CR, and UKAEA, the Karlsruher Instit\"ut f\"ur Technologie (KIT) Order No. 320/20459037/INR-NK, and the CNCSIS-Bucharest under Contract PN-II-ID-PCE-2008-2-448.

\end{document}